\documentclass[a4paper,12pt]{article}
\pdfoutput=1 

\usepackage{jheppub} 
\usepackage[export]{adjustbox}
\usepackage[utf8]{inputenc}
\usepackage{multirow}
\usepackage{soul}

\usepackage{slashed}
\usepackage{bm}

\usepackage{fancyvrb}
\usepackage{fvextra}

\usepackage{comment}

\usepackage{tikz}
\usepackage{tkz-euclide}
\usetikzlibrary{shapes.misc}
\usetikzlibrary{decorations.pathmorphing}	
\tikzset{
    v/.style={decorate, decoration={snake, segment length=3mm, amplitude=0.75mm}, draw},
    f/.style={draw=black, postaction={decorate},
        decoration={markings,mark=at position .6 with {\arrow[very thick]{latex}}}},
    fb/.style={draw=black, postaction={decorate},
        decoration={markings,mark=at position .4 with {\arrowreversed[very thick]{latex}}}},
    fnar/.style={draw=black},
    g/.style={decorate, draw=black,
        decoration={coil,amplitude=3pt, segment length=3.5pt}},
    s/.style={dashed,draw=black, postaction={decorate},
        decoration={markings,mark=at position .55 with {\arrow[very thick]{latex}}}},
    sb/.style={dashed,draw=black, postaction={decorate},
        decoration={markings,mark=at position .55 with {\arrowreversed[draw=black,very thick]{latex}}}},
    snar/.style={dashed,draw=black,line width =1.25pt},
    cross/.style={cross out, draw=black, minimum size=2*(#1-\pgflinewidth), inner sep=0pt, outer sep=0pt},
cross/.default={3pt},
}
\usepackage{pifont}

\graphicspath{{Figures/}} 

\newcommand{\al}[1]{\begin{align}\begin{aligned} #1 \end{aligned}\end{align}}

\def\be{\begin{equation}}
\def\ee{\end{equation}}
\newcommand{\ba}{\begin{array}}
\newcommand{\ea}{\end{array}}
\def\bV{\begin{Verbatim}[breaklines=true, breakanywhere=true]}
\def\eV{\end{Verbatim}}

\title{\boldmath 
Dark Photon bounds in the dark EFT\\
}

\author[a,b]{Daniele Barducci,}
\author[c]{Enrico Bertuzzo,}
\author[d]{Giovanni Grilli di Cortona}
\author[c]{and Gabriel M. Salla}

\affiliation[a]{Universit\`{a} degli Studi di Roma la Sapienza, Piazzale Aldo Moro 5, 00185, Roma, Italy}
\affiliation[b]{INFN Section of Roma 1, Piazzale Aldo Moro 5, 00185, Roma, Italy}
\affiliation[c]{Instituto de F\'{i}sica, Universidade de S\~{a}o Paulo, C.P. 66.318, 05315-970 S\~{a}o Paulo, Brazil}
\affiliation[d]{Istituto Nazionale di Fisica Nucleare, Laboratori Nazionali di Frascati, C.P. 13, 00044 Frascati, Italy}

\emailAdd{daniele.barducci@roma1.infn.it}
\emailAdd{bertuzzo@if.usp.br}
\emailAdd{grillidc@lnf.infn.it}
\emailAdd{gabriel.massoni.salla@usp.br}

\abstract{Dark photons are massive abelian gauge bosons that interact with ordinary photons via a kinetic mixing with the hypercharge field strength tensor. 
This theory is probed by a variety of different experiments and limits are set on a combination of the dark photon mass and kinetic mixing parameter. These limits can however be strongly modified by the presence of additional heavy degrees of freedom.
Using the framework of dark effective field theory, we study how robust are the current experimental bounds when these new states are present. 
We focus in particular on the possible existence of a dark dipole interaction between the Standard Model leptons and the dark photon. 
We show that, under certain assumptions, the presence of a dark dipole modifies existing supernov\ae\ bounds for cut-off scales up to ${\cal O}(10-100\;{\rm TeV})$. On the other hand, terrestrial experiments, such as LSND and E137, can probe cut-off scales up to  ${\cal O}(3\;{\rm TeV})$. For the latter experiment we highlight that the bound may extend down to vanishing kinetic mixing.}

\begin{document} 
\maketitle

\section{Introduction}

The idea of the existence of dark sectors has become more and more attractive in the last few years. On the one side, the indisputable evidences for dark matter and dark energy make their existence at least plausible.
On the other hand, the null results from searches for electroweak (EW) and TeV scale New Physics (NP) at the LHC have given a strong push to the idea that NP may be secluded from the Standard Model (SM) particles. 
The simplest dark sector is perhaps the so-called dark photon model~\cite{Holdom:1985ag}, see~\cite{Fabbrichesi:2020wbt,Graham:2021ggy} for recent reviews, in which a new abelian gauge boson is added to the SM particle content. Its interactions with SM states arise solely from the kinetic mixing between the dark photon and the hypercharge gauge boson field strength tensors.
Such kinetic mixing is typically constrained to be small by a plethora of experimental data. Focusing on the so-called visible dark photon, {\emph{i.e.}} a dark photon with mass $m_{A'} > 1$ MeV with negligible decays into invisible states belonging to the dark sector, we can list accelerator experiments~\cite{Batell:2009di,Essig:2010gu,Bjorken:2009mm,Andreas:2012mt}, supernov\ae\ constraints~\cite{Dent:2012mx,Kazanas:2014mca,Rrapaj:2015wgs,Chang:2016ntp, Hardy:2016kme}, as well as bounds coming from Big Bang Nucleosynthesis (BBN) and Cosmic Microwave Background (CMB) data~\cite{Fradette:2014sza}. These bounds, which constrain the region with small kinetic mixing, are the ones that will be relevant for us in what follows. For additional bounds, we refer the reader to~\cite{Fabbrichesi:2020wbt,Lin:2019uvt,Ilten:2018crw}.

Once the SM is extended with a new abelian group it is easy to imagine that new beyond the SM (BSM) states, charged both under the SM and the new symmetry, might exist. This motivates us to consider and answer the following question:
\begin{quote}
\textit{how robust are the experimental bounds listed above if new states that interact both with the SM and the dark photon are present?}
\end{quote}
Given the vast number of possibilities for the presence of additional NP states and the limits from LHC direct searches for particles charged under the SM symmetries, we will assume for definiteness that such states have masses above the EW scale and we will use the Effective Field Theory (EFT) paradigm to frame our discussion. From a practical point of view, our approach amounts to adding non-renormalizable operators, suppressed by powers of the cutoff scale $\Lambda$ where these operators are generated, that connect the dark photon with the SM states. In order to avoid stringent bounds on $\Lambda$, present when the operators are generated by states charged under QCD, in the remainder of the paper we will only consider the phenomenology of operators that can be generated by uncolored states. More specifically, we will focus on the possible existence of a dark dipole moment for electrons which, below the EW scale, is of the form $A'_{\mu\nu} \bar{e}_L \sigma^{\mu\nu} e_R$ and discuss its phenomenological implications.~\footnote{See also~\cite{Dobrescu:2004wz} for a discussion of the dark dipole in the context of a massless dark photon, and~\cite{Rizzo:2021lob} for a different phenomenological considerations on dark dipole moments in the massive case.}

\begin{figure}[t]
\centering
\includegraphics[width=1.0\textwidth]{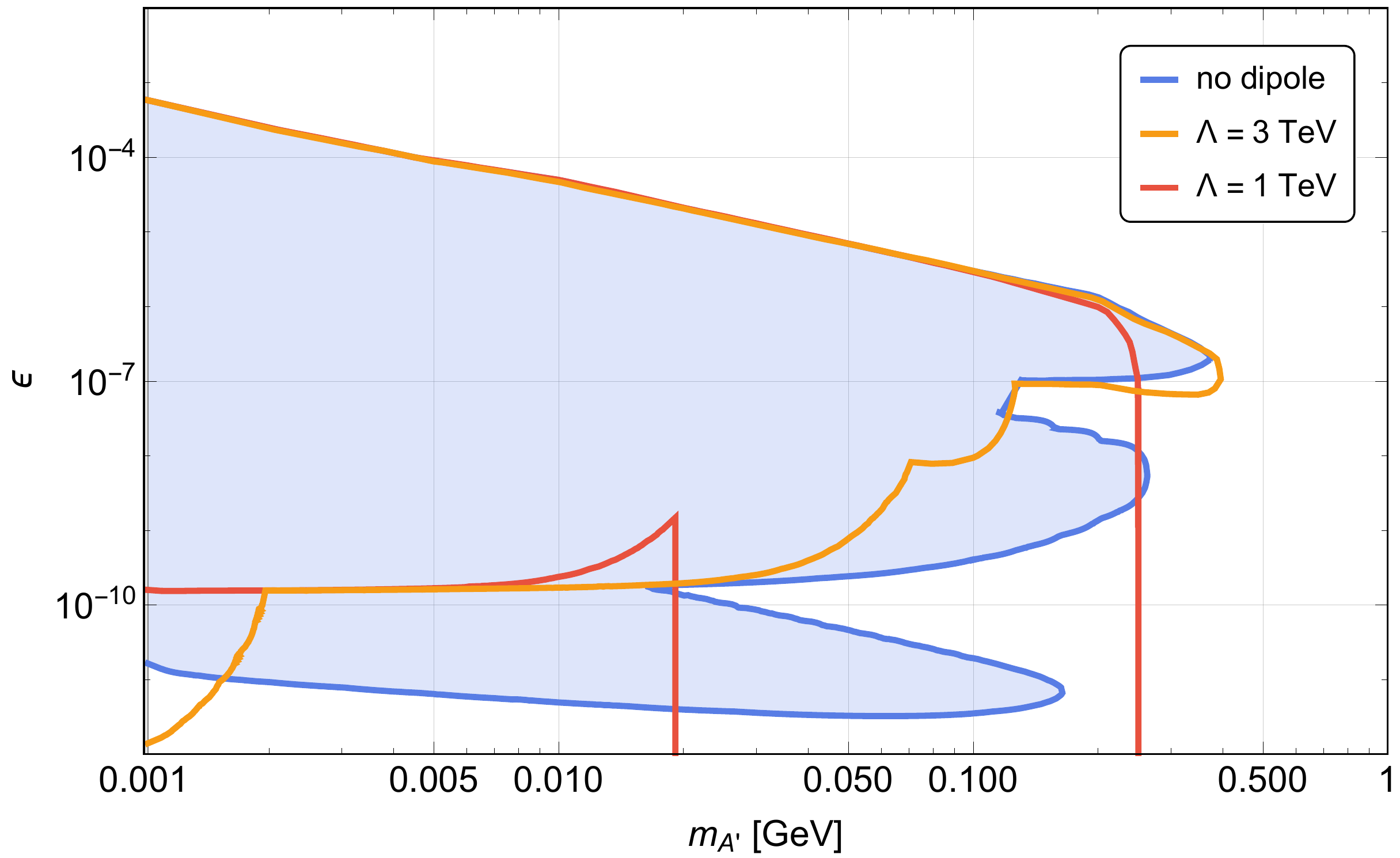}
\caption{Bounds for a dark photon model with and without a dark dipole interaction between the 
leptons and the dark photon. The bound is a combination of limits from SN1987A, LSND and E137, see the main text for more details.
}
\label{fig:all_bounds_intro}
\end{figure}

For convenience, we summarize in Fig.~\ref{fig:all_bounds_intro} the final results of this paper. The blue shaded area is the region excluded for a dark photon model without the dipole operator \cite{Batell:2009di,Essig:2010gu,Bjorken:2009mm,Andreas:2012mt,Dent:2012mx,Kazanas:2014mca,Rrapaj:2015wgs,Chang:2016ntp, Hardy:2016kme}.
From larger to smaller $\epsilon$, we show bounds coming from E137 (upper lobe), LSND (between the first and second lobe) and supernov\ae\ (the two lower lobes). In the presence of the dark dipole, the limits may change drastically. For a NP scale $\Lambda = 1$\;TeV we exclude the region inside the red curve, while for $\Lambda = 3$\; TeV the excluded region is inside the yellow curve. In the former case, the bounds from E137 increase dramatically, excluding $\epsilon\lesssim 10^{-6}$ for dark photon masses in the range $(20 \div 250)$\;MeV. In the latter case, the regions excluded by E137 and LSND slightly increase. In both cases, the supernov\ae\ bound becomes much weaker. Overall, the presence of the dipole operator can dramatically affect the dark photon bounds. The limits just described are derived fixing the dark electron dipole moment to unity. From the UV perspective, however, this choice may require some tuning of the parameters. We will get back to this issue in Sec.~\ref{sec:UV_compl} and in the Conclusions.

The paper is organized as follows. In Sec.~\ref{sec:framework} we fix our notation and describe the EFT that will be considered, by sketching also possible ultraviolet (UV) completions that lead to the configuration on which we will focus.
In Sec.~\ref{sec:dark_dipole} we perform the phenomenological analysis of the effects of dark dipole moments on the usual experimental bounds and present our findings, discussing in detail the features of Fig.~\ref{fig:all_bounds_intro}. We conclude in Sec.~\ref{sec:conclusion}. Finally, in App.~\ref{app:diagonalization} we present the derivation of the diagonalization of the dark photon kinetic and mass Lagrangian in the presence of higher dimensional operators.

\section{Theoretical framework}\label{sec:framework}

In this section we introduce the dark EFT operators and sketch a few UV completions that can generate a dark lepton dipole independently from the kinetic mixing, allowing us to consider both of them as free parameters of the effective theory.

\subsection{The Dark Photon Effective Field Theory}\label{sec:EFT}

Our goal is to analyze the phenomenological consequences of embedding the dark photon setup in an EFT. For concreteness, we will consider the SM extended by an abelian $U(1)_X$ group, under which all the SM states are uncharged. The gauge boson associated to this new symmetry will be denoted by $X_\mu$. At the $d\le 4$ level we consider the following Lagrangian
\be\label{eq:L_ren}
{\cal L}_{d \leq 4} = {\cal L}_{{\rm SM}} - \frac{1}{4\, \tilde{g}_d^2} X_{\mu\nu} X^{\mu\nu} + \frac{\kappa}{2\, \tilde{g}'\, \tilde{g}_d} B_{\mu\nu} X^{\mu\nu} +\frac{m_X^2}{2} X_\mu X^\mu  ,
\ee
where, at this stage, we are working in the basis in which the gauge couplings appear in the kinetic term and where we are agnostic about the origin of the mass for the dark photon $X_\mu$.\footnote{We can for example postulate it to have a St\"uckelberg mass~\cite{Stueckelberg:1938hvi}, {\emph{i.e.}} we can suppose that the scalar responsible for the $U(1)_X$ spontaneous breaking, if present, has a mass above the cutoff $\Lambda$.}
\begin{table}[t!]
\begin{center}
\begin{tabular}[t]{c|c}
\multicolumn{2}{c}{Dipole operators} \\
\hline \hline
  & Operator  \\
\hline
${\cal O}_{Xu}$ & $X_{\mu\nu} \bar{Q} \tilde{H} \sigma^{\mu\nu} u_R$    \\
${\cal O}_{Xd}$ & $X_{\mu\nu} \bar{Q} H \sigma^{\mu\nu} d_R$ \\
${\cal O}_{Xe}$ &  $X_{\mu\nu} \bar{L} H \sigma^{\mu\nu} e_R$ \\
${\cal O}_{Bu}$ & $B_{\mu\nu} \bar{Q} \tilde{H} \sigma^{\mu\nu} u_R$    \\
${\cal O}_{Bd}$ & $B_{\mu\nu} \bar{Q} H \sigma^{\mu\nu} d_R$  \\
${\cal O}_{Be}$ &  $B_{\mu\nu} \bar{L} H \sigma^{\mu\nu} e_R$  \\
${\cal O}_{Wu}$ & $W^a_{\mu\nu} \bar{Q} T^a_L \tilde{H} \sigma^{\mu\nu} u_R$     \\
${\cal O}_{Wd}$ & $W^a_{\mu\nu} \bar{Q} T^a_L H \sigma^{\mu\nu} d_R$  \\
${\cal O}_{We}$ &  $W^a_{\mu\nu} \bar{L}T^a_L H \sigma^{\mu\nu} e_R$  
\end{tabular}
\hfill
\begin{tabular}[t]{c|c}
\multicolumn{2}{c}{Operators generating kinetic terms or kinetic mixing} \\
\hline \hline 
  & Operator \\
\hline
${\cal O}_{XX}$      &    $H^\dag H X_{\mu\nu} X^{\mu\nu}$   \\
${\cal O}_{BB}$      &    $H^\dag H B_{\mu\nu} B^{\mu\nu}$   \\
${\cal O}_{WW}$      &    $H^\dag H W^a_{\mu\nu} W^{a\mu\nu}$   \\
${\cal O}_{XB}$      &    $H^\dag H X_{\mu\nu} B^{\mu\nu}$   \\
${\cal O}_{BW}$      &    $H^\dag T^a_L H W^a_{\mu\nu} B^{\mu\nu}$  \\
${\cal O}_{XW}$      &    $H^\dag T^a_L H W^a_{\mu\nu} X^{\mu\nu}$  \\
\multicolumn{2}{c}{Other relevant operators} \\
\hline \hline
  & Operator  \\
\hline
${\cal O}_{T}$      &    $\left|H^\dag D_\mu H \right|^2$   \\
\end{tabular}
\caption{Operators considered in this work. We use the standard notation for the SM states, while the gauge boson associated with the $U(1)_X$ symmetry is denoted by $X$. The symbol $T^a_L$ denotes the $SU(2)_L$ generators normalized as Tr$[T_L^a,T_L^b]=\delta^{ab}/2$.}
    \label{tab:operators}
\end{center}
\end{table}
In addition to the Lagrangian of Eq.~\eqref{eq:L_ren}, we also consider the effective interactions described by
\be\label{eq:L_EFT}
{\cal L}_{d=6} = \frac{1}{16\pi^2}\frac{1}{\Lambda^2} \sum_{i \neq T} {\cal C}_i \,{\cal O}_i + \frac{1}{\Lambda^2}\mathcal{C}_T \,{\cal O}_T \, ,
\ee
where the set of operators considered is presented in Tab.~\ref{tab:operators}. These operators can be classified in {\emph{i)}} dipole operators, {\emph{ii)}} operators that modify the kinetic terms or that contribute to the vector bosons kinetic mixing and {\emph{iii)}} the operator ${\cal O}_T$ that contributes to the $Z$ boson mass.\footnote{We do not consider possible CP-violating effects arising, {\emph{e.g.}} from operators involving dual field strength tensors.} With the exception of ${\cal O}_T$, all these operators must be generated at loop level in a weakly coupled UV completion~\cite{Buchmuller:1985jz,Craig:2019wmo}, thus justifying the introduction of the factor $1/16\pi^2$ in Eq.~\eqref{eq:L_EFT}. The ${\cal C}_i$ Wilson coefficients associated with the dipole operators are, a priori, $3\times 3$ complex matrices in flavor space. For simplicity, we will consider only the dark dipole associated with the electron. We call the Wilson coefficient in the lepton sector $d_e$. 

Following the steps outlined in App.~\ref{app:diagonalization} we obtain the following interaction Lagrangian in the EW broken phase
\al{\label{eq:Lint}
{\cal L}_{\rm int} & = e A_\mu J_Q^\mu +e \left( c_W \epsilon - s_W {\cal S}_d \right) J_Q^\mu A_\mu ' \\
& + \left(\sqrt{g^2 + g'^2} \left(1 + s_W c_W {\cal S} \right) \left[J_3^\mu - s_W^2 J_Q^\mu \right] - e (c_W^2 - s_W^2) {\cal S} J_Q^\mu \right) Z_\mu \\
& + \frac{d_e}{16\pi^2} \frac{v}{\Lambda^2} \bar{e}_L \sigma^{\mu\nu} e_R A_{\mu\nu}' + h.c. \, .
}
In the previous expression $A_\mu$, $Z_\mu$ and $A^\prime_\mu$ describe the physical photon, $Z$ boson and dark photon respectively, $c_W$ and $s_W$ are the usual cosine and sine of the weak angle, $g$ and $g^\prime$ the coupling constants of the SM $SU(2)_L$ and $U(1)_Y$ gauge groups,
$\epsilon$ is the kinetic mixing once the effect of higher dimensional operators is taken into account, see Eq.~\eqref{eq:epsilon}, and the dark electron dipole moment $d_e$ is a linear combination of the dipole operators Wilson coefficients. We focus on dipole operators involving electrons since, as already mentioned in the Introduction, dipole operators involving quarks will in general be generated by colored heavy states on which the bounds from direct searches are stronger. Finally ${\cal S}$ and ${\cal S}_d$, defined in Eq.~\eqref{eq:S_Sd}, describe the $W^3_{\mu\nu} B^{\mu\nu}$ and $W^3_{\mu\nu} X^{\mu\nu}$ kinetic mixings respectively. Apart from a normalization, ${\cal S}$ is the usual Peskin-Takeuchi parameter~\cite{Peskin:1990zt,Peskin:1991sw}, while ${\cal S}_d$ is an analogous parameter involving the dark photon. As expected, the higher dimensional operators modify in the usual way the coupling of the $Z$ boson, while the kinetic mixing introduces the usual dark photon coupling to the electromagnetic current. We have nevertheless a novel effect: an additional term in the coupling between the dark photon $A'$ and the electromagnetic current due to the $W^3_{\mu\nu} X^{\mu\nu}$ operator, independent of the usual kinetic mixing.

As we are going to see in Sec.~\ref{sec:UV_compl}, it is possible to have UV completions in which ${\cal S}_d$ is subdominant with respect to $\epsilon$, and the dark lepton dipole $d_e$ is independent of the kinetic mixing. Moreover, we will assume that the electron dark dipole can be taken to be of order unity although, as we will see, this may require some tuning.
To simplify the  phenomenological analysis of Sec.~\ref{sec:dark_dipole} we will then make the following assumptions: {\emph{i)}} we take  the parameters $\left\{\epsilon,d_e \right\}$ as independent and neglect the contribution from ${\cal S}_d$, see Sec.~\ref{sec:UV_compl}; {\emph{ii)}} we assume $d_e$ to be real to avoid the stringent bounds from electron dipole moment measurements~\cite{Panico:2018hal}. Altogether we thus analyze, under these assumptions, the phenomenology of the following interaction Lagrangian
\be\label{eq:Lint2}
{\cal L}_{\rm int} = e\,c_W\,\epsilon A_\mu' J_Q^\mu + \frac{d_e}{16\pi^2} \frac{v}{\Lambda^2} \bar{e}_L \sigma^{\mu\nu} e_R A_{\mu\nu}^\prime + h.c. \, .
\ee

\subsection{A sketch of possible UV completions}\label{sec:UV_compl}

We now present a few examples to illustrate how the Lagrangian in Eq.~\eqref{eq:Lint2} may be generated. Our purpose is to show that our assumptions are justified, {\emph{i.e.}} that we can have ${\cal S}_d$ negligible with respect to $\epsilon$ and take $\epsilon$ and $d_e$ to be independent parameters.

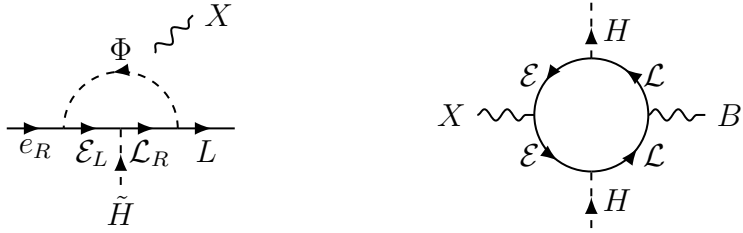
\begin{figure}[tb]
\begin{center}
\adjustbox{valign=m}{
\begin{tikzpicture}[line width=0.75] 
\draw[f] (-0.75, 0) -- (0,0)node[midway, below]{$e_R$};
\draw[f] (0,0) -- (0.75, 0)node[midway, below]{${\cal E}_L$};
\draw[f] (0.75,0) -- (1.5,0)node[midway,below]{${\cal L}_R$};
\draw[f] (1.5,0)--(1.5+0.75,0)node[midway, below]{$L$};
\draw[s] (0.75,-0.75) node[below]{$\tilde{H}$}-- (0.75,0) ;
\draw[s] (1.5,0)  arc (0:180:0.75) node[midway,above]{$\Phi$};
\draw[v] (1.2,1) -- (1.7,1.5) node[right]{$X$};
\end{tikzpicture}}
\hspace{2.0cm}
\adjustbox{valign=m}{
\begin{tikzpicture}[line width=0.75] 
\draw[v] (-0.75, 0) node[left]{$X$}-- (0,0);
\draw[f] (0.75,0.75) arc (90:180:0.75) node[midway, left]{${\cal E}$};
\draw[f] (1.5,0) arc (0:90:0.75)node[midway,right]{${\cal L}$};
\draw[s] (0.75,0.75) -- (0.75, 1.5)node[midway,right]{$H$};
\draw[v] (1.5,0) -- (1.5+0.75,0)node[right]{$B$};
\draw[f] (0, 0) arc (180:270:0.75)node[midway,left]{${\cal E}$};
\draw[f] (0.75,-0.75) arc (-90:0:0.75)node[midway,right]{${\cal L}$};
\draw[s] (0.75, -1.5) -- (0.75, -0.75)node[midway,right]{$H$};
\end{tikzpicture}} 
\end{center}
\caption{\label{fig:diagrams1} Representative diagrams leading to the generation of the kinetic mixing and dark dipole moments at the one-loop level in the model of Eq.~\eqref{eq:LVLF}. See the main text for more details.}
\end{figure}
Let us consider a framework in which a set of vector-like leptons is added to the SM particle content. These fermions, denoted by ${\cal L}$ and ${\cal E}$, have the same gauge quantum numbers as the SM lepton left-handed doublet and right-handed singlet respectively and we assign to them a charge 
$q$ under the $U(1)_X$ symmetry. {Moreover, we assume that another symmetry forces the heavy fermions to couple only to the first generation. This may be a $\mathbb{Z}_2$ symmetry under which only the heavy fermions and the leptons of the first generation are charged, but the exact nature of such symmetry will not be important for our discussion.} We further add a scalar $\Phi$, complete singlet under the SM gauge group but carrying charge $q$ under $U(1)_X$. To avoid the generation of a large mass for the dark photon we assume $\langle \Phi \rangle = 0$. This particle content allows to write the following UV Lagrangian
\be\label{eq:LVLF}
{\cal L}_{{\rm UV}} = \kappa_L\,  \Phi^\dag \bar{L} {\cal L}_R + \kappa_E\, \Phi^\dag \bar{e}_R {\cal E}_L +y  \bar{\cal L}_L H {\cal E}_R + y' \bar{\cal L}_R H {\cal E}_L +  h.c. \ .
\ee
It is easy to see that these interactions generate, at one-loop level, all the relevant operators listed in Tab.~\ref{tab:operators} (see Fig.~\ref{fig:diagrams1} for some representative diagrams). First of all, the Wilson coefficient associated with the dark dipole results in
\be\label{eq:dark_dipole_WC}
{c_{Xe} = - \frac{\kappa_L \kappa_E^* \, g_X\, q\, (2 y - y')}{192 \pi^2 \, M_{\rm NP}^2}} .
\ee
{In addition, the} usual $S$, $T$, $W$ and $Y$ parameters, generated respectively by ${\cal O}_{BW}$, ${\cal O}_{T}$, ${\cal O}_{WW}$ and ${\cal O}_{BB}$, give a lower bound of about $400$ GeV on the heavy fermion and scalar masses, assumed degenerate for simplicity,\footnote{Explicit expressions for the electroweak parameters generated by vector-like fermions can be found in~\cite{Angelescu:2020yzf}.} while direct searches at colliders put a bound of about $800$ GeV on the masses of the new states~\cite{Guedes:2021oqx}. This bound, however, strongly depends on the invisible branching ratio of the vector-like fermions. More importantly, the operators ${\cal O}_{XB}$ and ${\cal O}_{XW}$ are also generated with Wilson coefficients
\be\label{eq:Wilson1}
c_{XB} =  \frac{g'\, g_X \, q\, (y^* y' + y y'^*)}{16\pi^2 \, M_{\rm NP}^2}\, , ~~~ c_{XW} = - \frac{13\, g\, g_d\, q\, |y|^2}{480\pi^2\, M_{\rm NP}^2}\, ,
\ee
where we indicate with $M_{\rm NP}$ the common mass for the NP states and $g_d$ is the gauge coupling of the $U(1)_X$ symmetry.
The first operator contributes to $\epsilon$, see Eq.~\eqref{eq:epsilon}, while the second one generated the ${\cal S}_d$ parameter appearing in Eq.~\eqref{eq:Lint}. If the value of the original kinetic mixing parameter $\kappa$ is sufficiently suppressed and all couplings in ${\cal L}_{\rm UV}$ are of the same size, we obtain a kinetic mixing $\epsilon \sim c_{XB} v^2/(16\pi^2 \Lambda^2)$ of the same order of magnitude as the dark dipole ${d_e} v^2/(16\pi^2 \, \Lambda^2)$.
When this happens, the effect of the dark dipole is typically too small to be observed. It is nevertheless easy to imagine extensions of this framework in which the contributions to $c_{XB}$ and $c_{XW}$ are suppressed. This can be achieved by enriching the NP sector with a set of $N_s$ new scalar fields $\phi$ with quantum numbers under $(\bm{1}, \bm{2}, y_S, q_S)$ under ${\cal G}_{\rm SM}\times U(1)_X$, which generate
\be\label{eq:Wilson2}
c_{XB} = - N_s \frac{g' \, g_X\, y_S \, q_S\, \lambda}{48\pi^2\, M_{\rm NP}^2 }, ~~~ c_{XW} = - N_s \frac{g\, g_X\,  q_S\, \lambda}{48\pi^2\, M_{\rm NP}^2 },
\ee
where $\lambda$ is the quartic coupling $\lambda H^\dag H \phi^\dag \phi$, assumed to be universal among the $N_s$ copies of scalar doublets with mass $M_{\rm NP}$. Adding Eqs.~\eqref{eq:Wilson1} and~\eqref{eq:Wilson2}, we see that there are choices of parameters for which a cancellation happens and it is possible to strongly suppress both $c_{XB}$ and $c_{XW}$. If this is the case, in the Lagrangian of Eq.~\eqref{eq:Lint2} we will have $\epsilon\sim \kappa$, with the dark dipole ${d_e}$ uncorrelated with the kinetic mixing. {Another source of tuning may come from the electron Yukawa coupling. Indeed, the left diagram of Fig.~\ref{fig:diagrams1}, without the $X$ boson, generates the following 1-loop contribution to the electron Yukawa coupling $y_e$:}
\be
{\delta y_e = \frac{\kappa_L \kappa_E^* (y+y')}{32\pi^2} .}
\ee
{To ensure $d_e \gg y_e$ we either need $y' \simeq -y$ (which suppressed the electron Yukawa without suppressing the dark dipole of Eq.~\eqref{eq:dark_dipole_WC}) or we need to tune away this contribution against the bare electron Yukawa. For our phenomenological analysis we will always take $d_e \gg y_e$. We will comment further on this issue in the Conclusions.}

{The conditions} described above ensure that the dipole may dominate over the pure kinetic mixing, a situation we will {consider} in our phenomenological study of Sec.~\ref{sec:dark_dipole}. In what follows, we will remain agnostic about the UV origin of the couplings in Eq.~\eqref{eq:Lint2}, simply assuming $\epsilon$ and ${d_e}$ to be independent parameters.

\section{Phenomenology of a dark electron dipole moment}\label{sec:dark_dipole}

We now turn to the study of how the bounds on the dark photon parameter space are modified by the effect of the dark {electron} dipole of Eq.~\eqref{eq:Lint2}. Inspired by the considerations of Sec.~\ref{sec:UV_compl}, we will consider values of $\Lambda \gtrsim 0.5\;$TeV, {\it i.e.} satisfying the irreducible bound from electroweak precision measurements outlined in the previous section. As pointed out there, in specific UV completions the lower bound on $\Lambda$ may be set by direct searches. These bounds are however model dependent and will be ignored from now on.  

Depending on the process considered, the dark dipole can {\emph{i)}} affect both the dark photon production and decay or {\emph{ii)}} solely the decay. We will consider physical processes in which both situations can occur. We can identify several regions in parameter space in which different kinds of experiments are relevant, see {\emph{e.g}} Fig.~1 in~\cite{Fradette:2014sza}. For $\epsilon \gtrsim 10^{-7}$ the most relevant bounds come from accelerator experiments, for $10^{-10} \lesssim \epsilon \lesssim 10^{-(6 \div 7)} $ the most relevant bound comes from supernov\ae\ and, finally, for $\epsilon \lesssim 10^{-10}$ the relevant bounds come from BBN and CMB. We will discuss these bounds in turn, starting from the smallest values of $\epsilon$ and working our way up to larger values. A summary of our findings is shown in Fig.~\ref{fig:all_bounds_intro}.

As we will see, a prominent role in the modifications of the dark photon bounds is due to the additional contributions to the dark photon decay width due to the dark dipole. In particular, the dark photon partial width into a pair of {electrons} is modified as
\al{
\Gamma(A' \to {e^+ e^-}) & = \bigg[\frac{\alpha_{EM} \, c_W^2\, \epsilon^2}{3} m_{A'} \left(1+ 2 \frac{m_e^2}{m_{A'}^2} \right)+ \frac{\alpha_{EM}^{1/2} c_W \epsilon\, d_e}{8\pi^{5/2}} \frac{m_e\, m_{A'} \, v}{\Lambda^2}\\
& \qquad\qquad {} +\frac{d_e^2}{1536 \pi^5}m_{A'}^3  \frac{v^2}{\Lambda^4}  \left( 1+ 8 \frac{m_e^2}{m_{A'}^2}  \right) \bigg] \sqrt{1-\frac{4 m_e^2}{m_{A'}^2}}  \, .\label{eq:width}
}
The total dark photon width thus reads
\be\label{eq:total_width}
\Gamma_{A'} = \Gamma(A' \to e^+ e^-) + \Gamma(A' \to \mu^+ \mu^-) \left[1+R(m_{A'}) \right],
\ee
where $R = \sigma(e^+ e^- \to {\rm hadrons})/\sigma(e^+ e^- \to \mu^+ \mu^-)$ allows to include the decay into hadrons, see {\emph{e.g.}}~\cite{Ilten:2018crw}. It is important to notice that the term proportional to {$d_e^2$} scales as $m_{A'}^3$ and becomes more important for larger values of the dark photon mass. This effect will be important in what follows.

\subsection{Very dark photons and the bounds from BBN and CMB}\label{sec:BBN_CMB}
\begin{figure}
\begin{center}
\includegraphics[width=.49\textwidth]{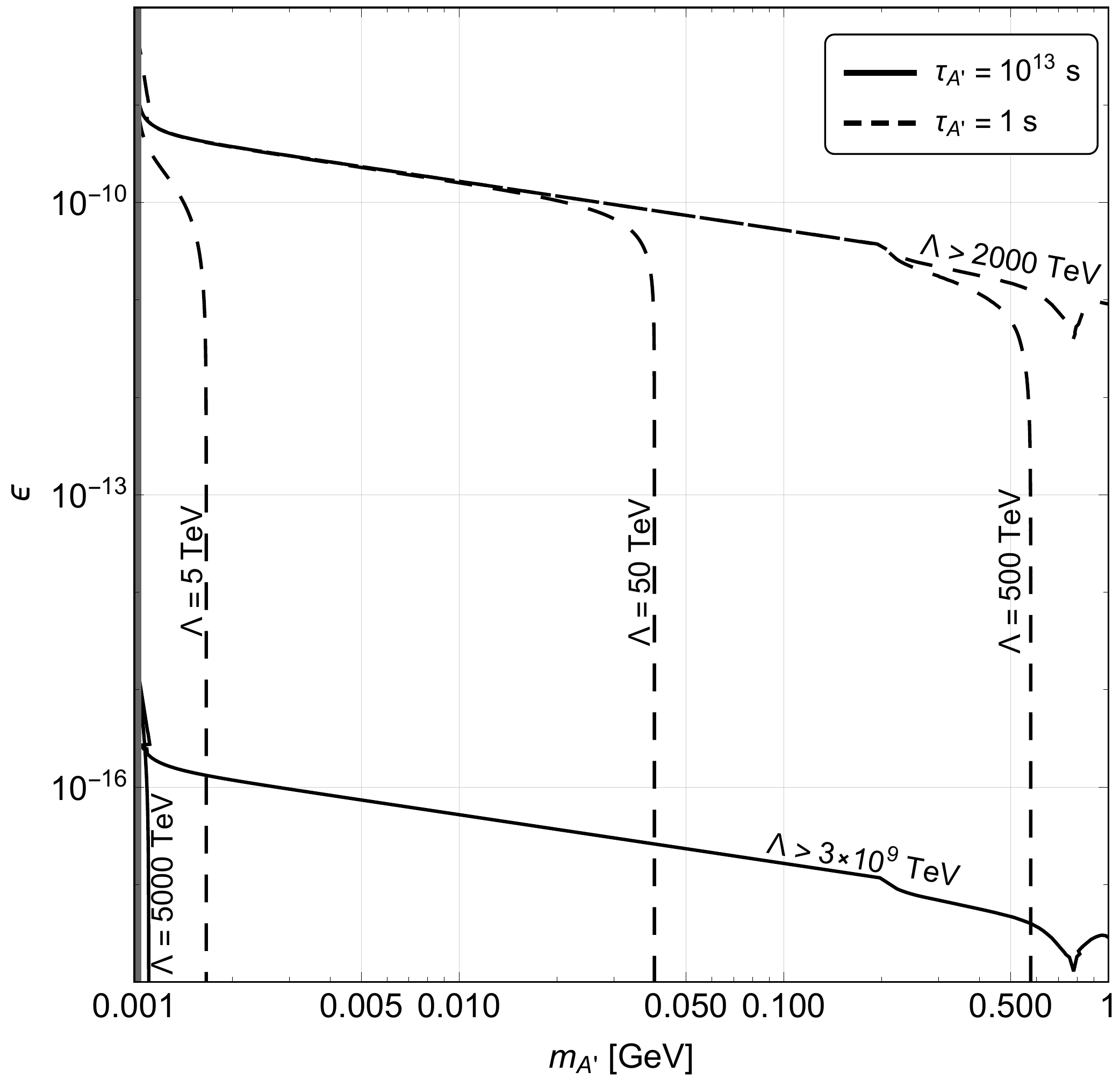}\hfill
\includegraphics[width=.48\textwidth]{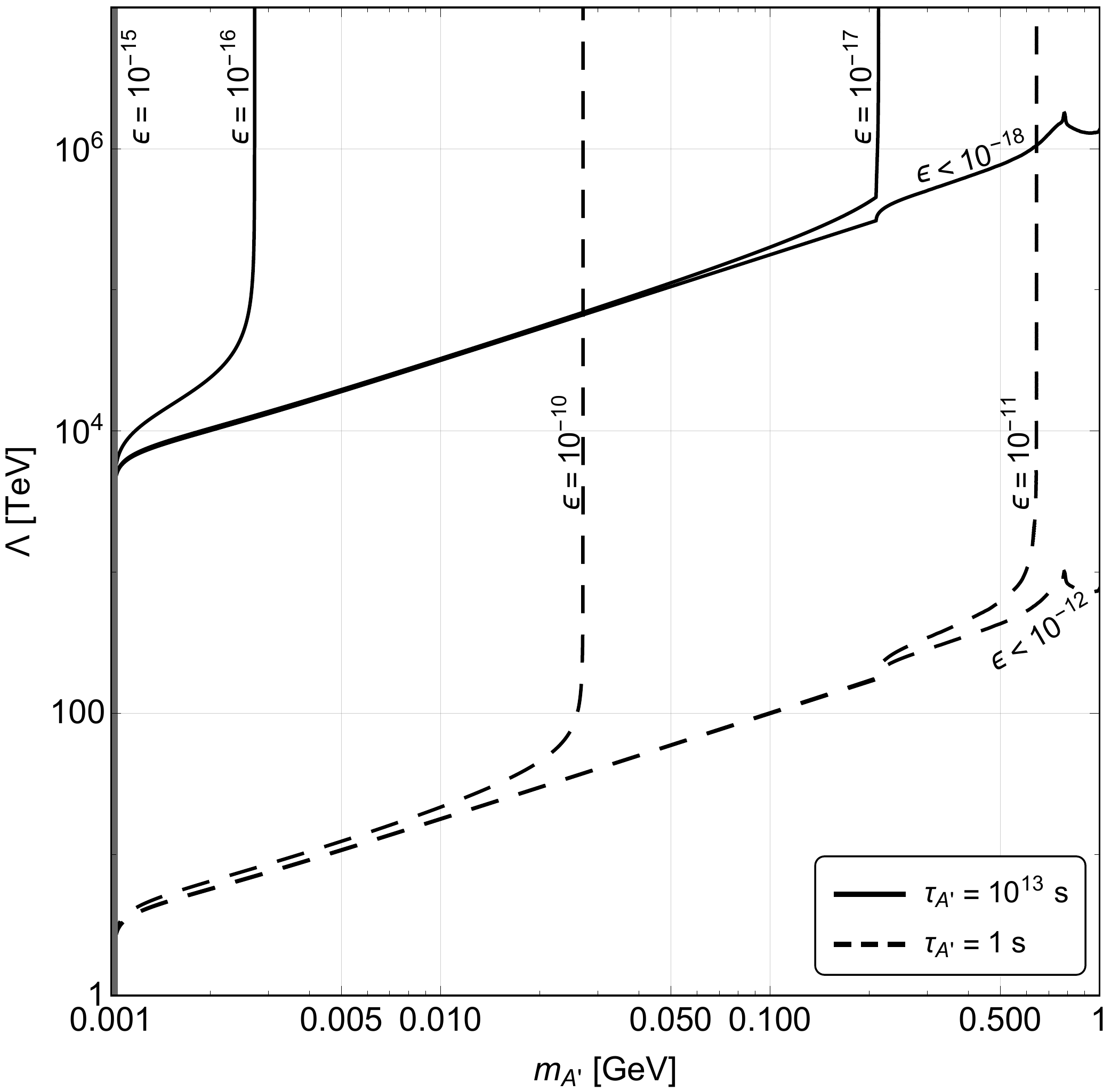} \\
\end{center}
\caption{\label{fig:lifetime} Contours of dark photon constant proper lifetime for different values of the dark dipole scale $\Lambda$ in the $m_{A^\prime}-\epsilon$ plane (left) or for different values of the mixing parameter $\epsilon$ in the $m_{A^\prime}-\Lambda$ plane (right). In both cases we fix ${d_e}=1$.}
\end{figure}
Let us start with the bounds on {\emph{very dark}} photons coming from BBN and CMB. The term {\emph{very dark}} photon refers to the case of very small kinetic mixing, $\epsilon \lesssim 10^{-10}$, where the dark photon has a lifetime larger than $1$\;s. Late decays with lifetimes larger than $1$\;s ($10^{13}$\;s) can inject electromagnetic energy that would interfere with the outcome of BBN (CMB).

We show in the left panel of Fig.~\ref{fig:lifetime} the dark photon lifetime, computed according to Eq.~\eqref{eq:total_width}, in the $(m_{A'}, \epsilon)$ plane for different values of the dark dipole scale $\Lambda$. The case without dipole is the limit for very large $\Lambda$.  When the dipole is turned on, we fix {$d_e = 1$} and give only the value of the cutoff scale $\Lambda$. The dashed lines in both panels show the contours of $\tau_{A'} = 1$\;s, while the solid ones show $\tau_{A'} = 10^{13}$\;s. The figure shows that the dipole operator {may modify} the dark photon lifetime dramatically. For a scale $\Lambda\sim 5$\;TeV the dark photon lifetime is below $1$\;s in most of the parameter space and it can never be $\mathcal{O}(10^{13})$\;s. The dark dipole contribution dominates for small values of $\epsilon$ and of $\Lambda$. It becomes negligible only for very large values of the dark dipole scale.  
The right panel of Fig.~\ref{fig:lifetime} illustrates instead the dark photon lifetime in the $(m_{A'}, \Lambda)$ plane for different values of the dark photon mixing $\epsilon$. It shows that the dark photon lifetime is larger than $1$\;s only for large values of $\Lambda$ and that dark photons decaying after the CMB era requires huge values of $\Lambda$. 

We conclude that even for rather heavy NP, the bounds coming from BBN and CMB are expected to be ineffective simply because the dark photon had already decayed by the time BBN and recombination happened. For larger values of $\Lambda$ some of the bounds may re-emerge. It is however beyond the scope of this paper to perform the computation for such heavy NP.

\subsection{Supernov{\ae} bounds}
In a supernova explosion most of the energy expelled, $\sim 90\%$ of the difference in the gravitational binding energy between the progenitor and the remnant star, leaves the collapsing star in the form of neutrinos~\cite{Raffelt:1996wa}. However, new hypothetical light particles can be produced in large numbers and decay after macroscopic distances, transporting energy away from the star. This mechanism competes with the energy loss due to the SM neutrinos. Such an anomalous energy transport is strongly constrained because the observed neutrino burst in association with the explosion of the supernova SN1987A agrees with the predictions based on SM simulations~\cite{Burrows:1986me,Burrows:1987zz}. As a consequence, light beyond the SM states with masses below the supernova temperature of tens of MeV must be very weakly coupled, in order to suppress their production inside the supernova. Alternatively, they must be coupled strongly enough such that they interact with SM states with a high probability, thus being reprocessed inside the star and not transporting away energy.

The dominant dark photon production in the supernova core is given by nucleon Bremsstrahlung scattering, with the dark photon coupled to the proton through the kinetic mixing operator. This production mechanism is not influenced by the presence of the dipole operator. In what follows we will analyse two mechanisms that potentially constrain dark photons exploiting the SN1987A data. 

The first argument assumes that the luminosity due to massive dark photon emission must not exceed the one due to neutrino emission.
The luminosity due to dark photon emissions $L_{A'}$ is given by the energy emission rate times the decaying factor integrated over the whole volume of the supernova core
\begin{equation}
L_{A'} = \left\langle Q_{A'}\,e^{-R_c/(c\tau_{A'})} \right\rangle\,V_c  < L_\nu \simeq 3\times 10^{53}\,\mathrm{erg}/\mathrm{s}\, ,
\label{eq:SN_lumi}
\end{equation}
where $\left\langle Q_{A'}e^{-R_c/(c\tau_{A'})} \right\rangle$ is the energy emission rate taken from \cite{Dent:2012mx,Kazanas:2014mca}, $R_c\sim10$ km is the core radius, $V_c=4/3\,\pi R_c^3$ and $c\, \tau_{A'}=c/\Gamma_{A'}$ is the decay length including the relativistic dilation factor $m_{A'}/E_{A'}$. 

Furthermore, if the dark photons decay inside the mantle, the resulting electron and positrons may eject material beyond the decay radius and start light emission earlier than the observed three hours delay. A bound on $\epsilon$ and $m_{A'}$ can be set requiring that the energy deposited by the decaying dark photons does not exceed the gravitational binding energy of the mantle. This condition is given by~\cite{Kazanas:2014mca}
\begin{equation}
\left\langle Q_{A'}(e^{-0.8 \frac{R_*}{c\tau_{A'}}}-e^{- \frac{R_*}{c\tau_{A'}} })\right\rangle V_* \Delta t < \frac{G_N M_* \delta M}{0.8 R_*}\, ,
\label{eq:SN_mantle}
\end{equation}
where the difference of the two exponential accounts for the dark photons decaying within $(0.8 R_*,\, R_*)$, $\Delta t \sim 1\;$s is the emission duration, $G_N$ is the gravitational constant, $M_*\simeq10 M_{\odot}$ is the star mass, with $M_\odot$ the solar mass, $\delta M\sim 0.1 M_\odot$ is the mantle mass and $R_*\sim 3\times 10^7$ km is the radius of the progenitor star.

\begin{figure}[t]
\centering
\includegraphics[width=1.0\textwidth]{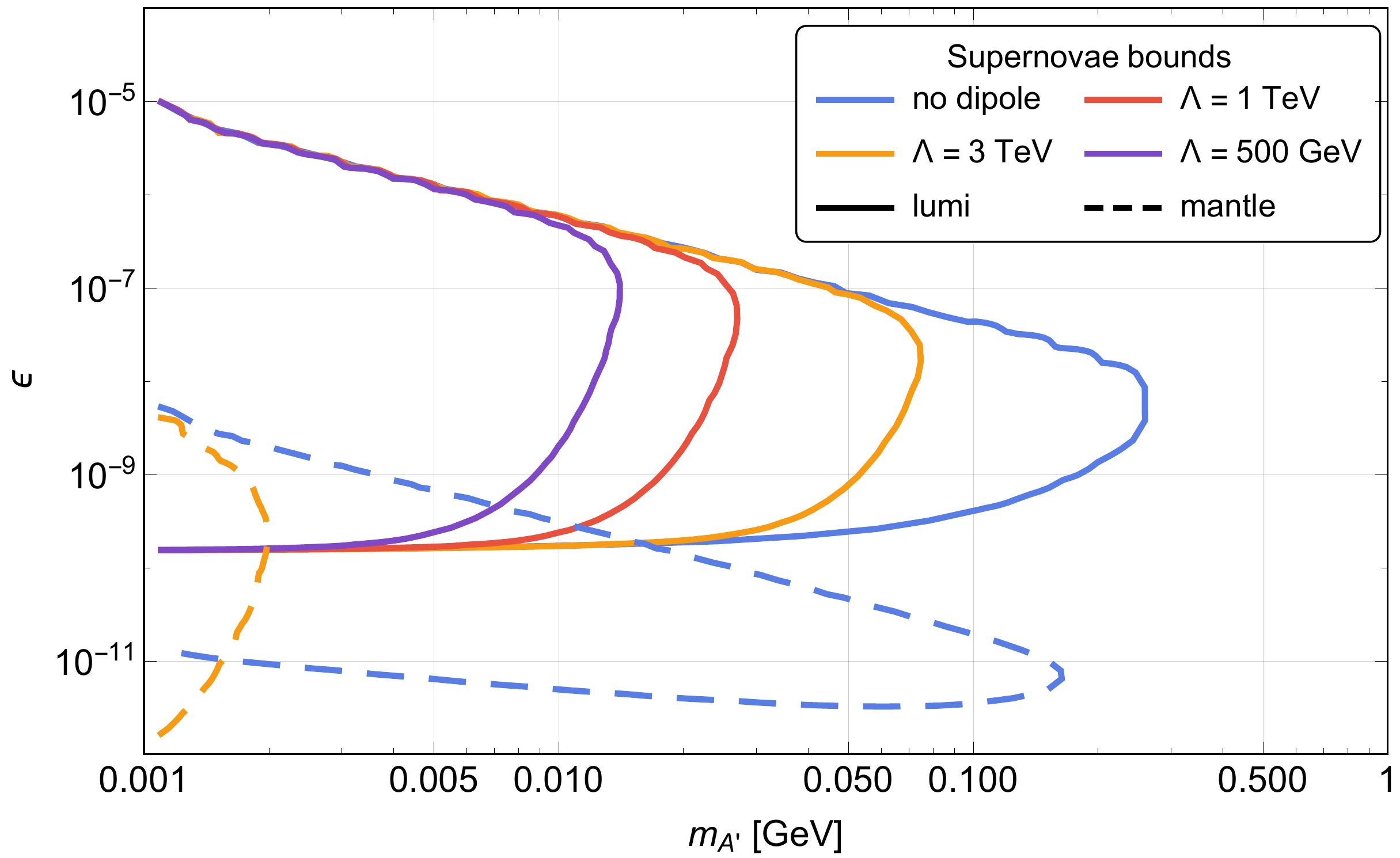}
\caption{Excluded region obtained for SN1987A. 
The solid curves are excluded by requiring that the dark photon luminosity is smaller than the neutrino one. The dashed curves are excluded by the dark photon decaying in the mantle argument.
The blue curves denote the limits obtained in the absence of the dipole operator, while the other curves show the limits for different values of $\Lambda$ by fixing ${d_e}=1$.}
\label{fig:SN_plot}
\end{figure}

Fig.~\ref{fig:SN_plot} shows the excluded region obtained by requiring the total energy emission rate due to dark photons to be smaller than the luminosity due to neutrino emission (solid curves) and the constraints coming from the decay inside the mantle argument (dashed curves). 
The solid blue curve describes the bound from a dark photon model without the dipole operator, in agreement with~\cite{Kazanas:2014mca}. The other solid curves, on the other hand, show the constraints for models including the dipole operator with {$d_e=1$} and $\Lambda = 0.5$ TeV (purple), $1$ TeV  (red) and $3$ TeV (yellow). The smaller the scale of the dipole operator $\Lambda$, the bigger is the effect on the decay constant and therefore the difference in the bound with respect to the model without the dipole operator. The inclusion of the dipole operator adds to the dark photon decay width a term that does not depend on $\epsilon$. Since this term increases the decay width, the exponential factor in equation \eqref{eq:SN_lumi} becomes more suppressed, leading to weaker bounds. 
The dashed blue curve shows the constraint from the argument of the dark photon decaying inside the mantle for a dark photon model without the dipole operator. In this case, the exponential factor in equation \eqref{eq:SN_mantle} is even more suppressed due to the presence in the exponent of the radius~$R_*$ of the progenitor star. As a consequence, already for $\Lambda=3$ TeV (dashed yellow curve) the bound is much weaker and, in the mass range of interest, is inexistent for smaller values of $\Lambda$.

We do not present bounds from the decay near the outside of the progenitor surface, because it would need a full fledged Monte Carlo simulation in order to estimate the annihilation cross section \cite{Kazanas:2014mca}. Furthermore, we did not include the effects of finite temperature and density on the kinetic mixing parameter $\epsilon$ \cite{Chang:2016ntp, Hardy:2016kme}, which are however expected not to change drastically our results for most of the parameter space of interest. These computations are out of the scope of this paper and will be presented elsewhere.

\subsection{Terrestrial bounds}\label{sec:terrestrial_bounds}

The region $\epsilon \gtrsim 10^{-7}$ can be bounded by several terrestrial experiments. Since the effect of the dark dipole moment is larger for lower values of $\epsilon$, and for TeV NP the effect of the dark {electron} dipole is practically irrelevant for $\epsilon \gtrsim 10^{-5}$, we will focus on the experiments that can probe the $10^{-7} \lesssim \epsilon \lesssim 10^{-5}$ region. By inspecting Fig.~1 of~\cite{Fradette:2014sza}, we will consider the neutrino experiment LSND~\cite{LSND:1996jxj} and the electron beam dump experiment E137~\cite{Bjorken:1988as}. For the former experiment, the dark {electron} dipole only affects the decay probability of the dark photon, while for the latter also the production rates get modified.

In full generality, the number of events inside the LSND or E137 detectors can be computed according to
\be\label{eq:Nevts}
N_{\rm evts} = N_{A'}\, f_{\rm geom} \, P_{\rm dec} \, ,
\ee
where $N_{A'}$ is the total number of dark photons produced, $f_{\rm geom}$ is the geometrical acceptance of the detector and $P_{\rm dec}$ is the probability for the observed decay $A' \to X$ to happen inside the detector. The latter quantity can be computed as
\be\label{eq:decay_prob}
P_{\rm dec} = e^{-\frac{d}{\lambda}}\Big(1-e^{-\frac{d+l}{\lambda}}\Big)\text{BR}\big(A'\rightarrow X\big)\, ,
\ee
where $d$ is the distance from the production point to the detector, $l$ the detector length and $\lambda$ is the dark photon decay length in the laboratory frame.

\subsubsection{LSND experiment }\label{sec:LSND}
\begin{figure}[tb]
\centering
\includegraphics[width=1.0\textwidth]{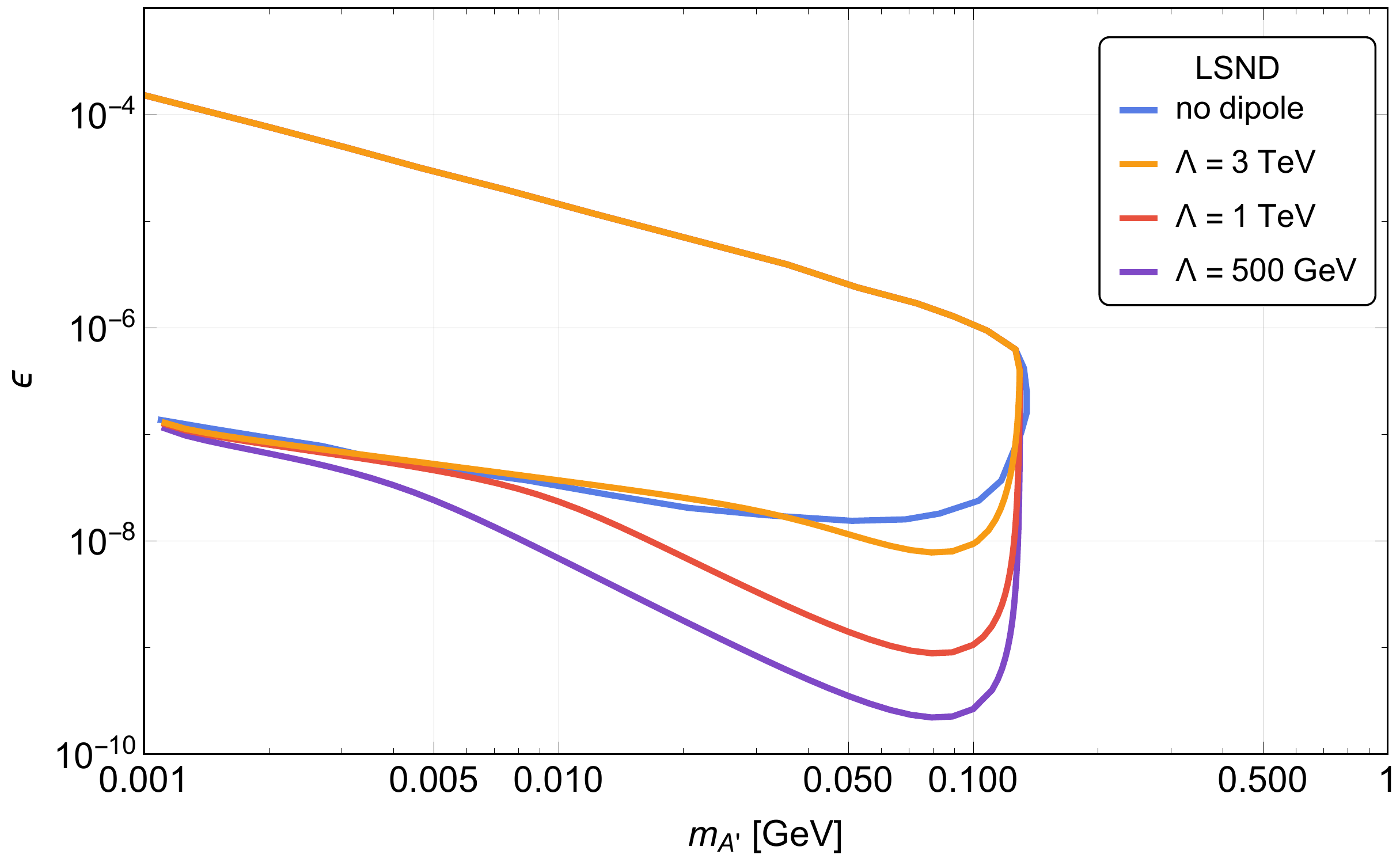}
\caption{Exclusion region obtained for LSND assuming the detection of at least 10 dark photon events.
The blue curve denotes the limit obtained in the absence of the dipole operator, while the other curves show the limits for different values of $\Lambda$ by fixing ${d_e}=1$.}
\label{fig:LSND_plot}
\end{figure}
The neutrino LSND experiment dumped around $10^{23}$ protons with an energy of $800\;$MeV on a water-copper target, producing a large number of pions that can decay into a dark photon-photon pair. Once produced, the dark photon can decay into an $e^+ e^-$ pair. If this happens inside the LSND detector, it may give rise to a signal that can be used to bound the dark photon parameter space. Such bounds have been discussed in~\cite{Batell:2009di, Essig:2010gu}. As pointed out there, there are considerable uncertainties, since no dedicated analysis was performed by the experimental collaboration for the signal under consideration and the bound relies on the misidentification of the $e^+e^-$ pair for a unique electron. In addition, uncertainties are also present in the determination of the pion flux~\cite{Batell:2009di}.

Given this situation, we will study the modifications of the bound induced by {$d_e \neq 0$} using the approximate procedure outlined in~\cite{Batell:2009di}. The number of produced dark photons is not affected by the presence of the dark dipole and is given by
\be
N_{A'} = N_{\pi^0} {\rm Br}(\pi^0\rightarrow \gamma A'), ~~~~~\text{Br}(\pi^0\rightarrow \gamma A')\simeq 2\epsilon^2\Big(1-\frac{m_{A'}^2}{m_{\pi^0}^2}\Big)^3,
\ee
where $N_{\pi^0}$ is the total number of neutral pions produced at the experiment. It can be reasonably approximated by the number of charged pions $N_{\pi^0}\simeq N_{\pi^\pm}\simeq 10^{21}$~\cite{Batell:2009di}. The geometrical acceptance $f_{\rm geom}$ in Eq.~\eqref{eq:Nevts} can be approximated as $
f_{\rm geom} \sim d\Omega/(4\pi)$, with $d\Omega$ the solid angle subtended by the detector. As for the decay probability, since we are interested in the modifications that take place in the region of small $\epsilon$, we can expand the exponential functions for $d, d+l \ll \lambda = \gamma c \tau$, approximating $\gamma \simeq m_{\pi^0}/m_{A'}$ and computing $\tau$ according to Eq. \eqref{eq:total_width}. For LSND we have $d=30$ m and $l=8.75$ m. This is an excellent approximation in the absence of dark dipole~\cite{Batell:2009di} and remains a very good one even when the dark dipole is switched on. Putting all together, our estimate for the number of events in the region of small $\epsilon$ is
\be\label{eq:Nevts_LSND}
N_\text{evts}\simeq \Phi_\nu V_\text{det}\frac{m_{A'}}{m_{\pi^0}}\text{BR}\big(\pi^0\rightarrow \gamma A'\big)\Gamma\big(A'\rightarrow e^+e^-\big),
\ee
where $\Phi_\nu = 1.3\times 10^{14}\, \nu/{\rm cm}^{-2}$ is the neutrino flux, $V_\text{det}=A_\text{det}l=2\times 10^8\text{ cm}^3$ is the volume of the detector and, as explained above, we consider only the decay into an $e^+e^-$ pair. The decay is, unlike production, affected by a non-vanishing dark dipole. The region in which $N_{\rm evts} > 10$ (which should reasonably give an indication of the excluded region, given the uncertainties mentioned above) is shown in Fig.~\ref{fig:LSND_plot}, fixing {$d_e = 1$} and choosing different values of $\Lambda$. We observe that, when {$d_e \neq 0$}, the exclusion is extended to lower values of $\epsilon$. The effect is larger for larger dark photon masses. This behaviour is expected since, for very small kinetic mixing, the contribution of the dipole in Eq.~\eqref{eq:width} (which grows with $m_{A'}^3$) dominates the decay width into electrons. In this region, the proper decay length $\lambda$ becomes smaller and the probability of decay $P_{\rm dec}$ in Eq.~\eqref{eq:Nevts} increases. As a consequence, the number of events increases and the excluded region becomes larger. Once more, we see that the effect is more evident for smaller values of $\Lambda$ and is quite small already for $\Lambda = 3$\;TeV.

\subsubsection{E137 experiment}\label{sec:E137}
\begin{figure}[tb]
\centering
\includegraphics[width=0.49\textwidth]{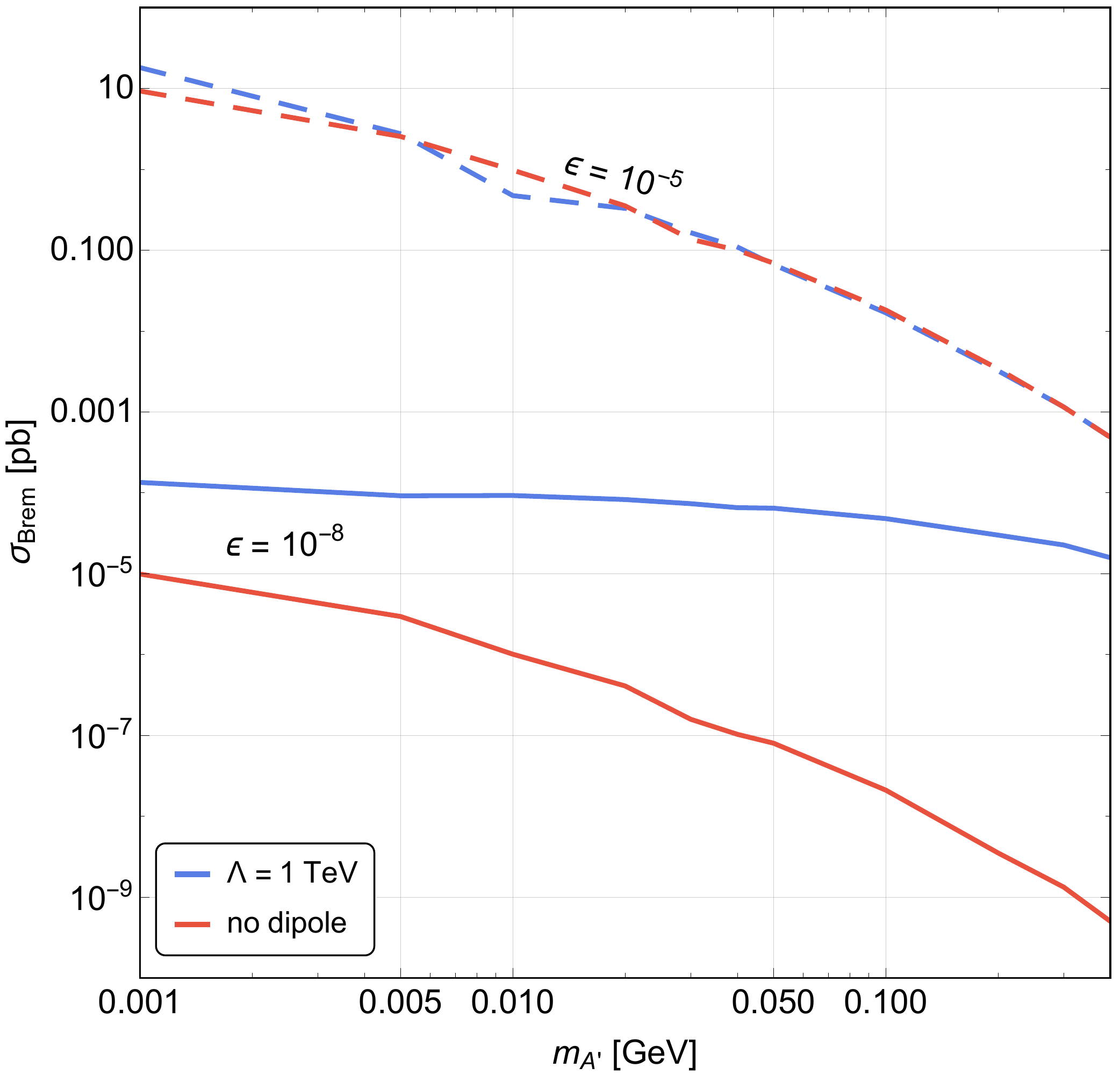}
\includegraphics[width=0.49\textwidth]{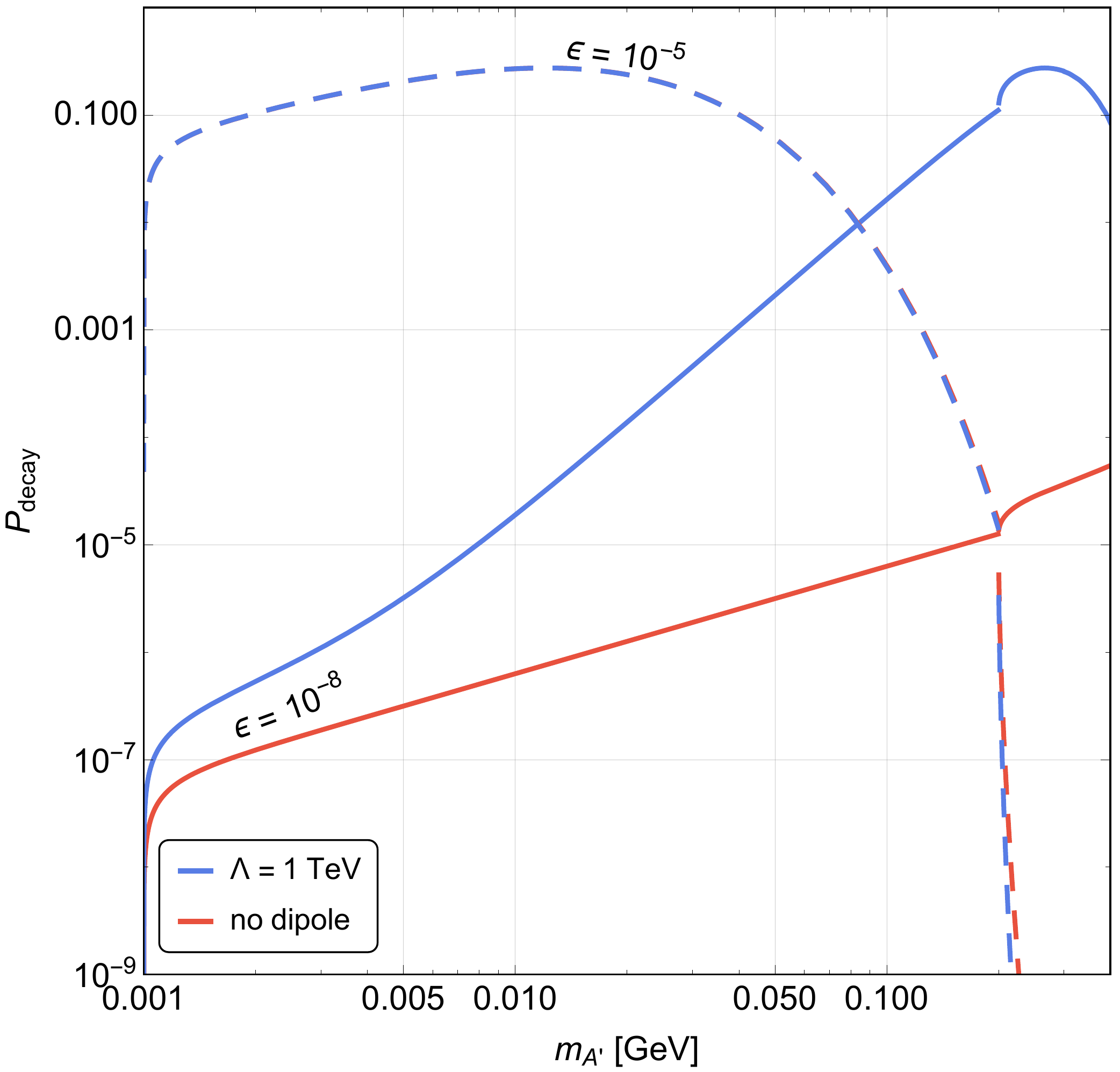} 
\caption{Left panel: Bremsstrahlung cross section as a function of the dark dipole mass for $\epsilon = 10^{-5}$ (dashed lines) and $\epsilon = 10^{-8}$ (solid lines) in the absence of the dark dipole (red lines) and with a dark dipole scale $\Lambda = 1$\;TeV (blue lines). Right panel: decay probability into leptons computed according to Eq.~\eqref{eq:decay_prob}. Same color code as in the left panel. We fix ${d_e}=1$.}
\label{fig:E137_xsec}
\end{figure}
The E137 experiment dumped 30\;C of electrons with energy of 20\;GeV on an aluminum target. The reaction products travelled through a hill (179 m) and a subsequent open region (204 m) before hitting the detector. Once the dark photon is produced via Bremsstrahlung off the electron beam, it can travel through the hill and the open space and decay into an $e^+e^-$ pair. Such decay can be detected and used to put bounds on the dark photon parameter space~\cite{Bjorken:2009mm,Andreas:2012mt}. Since both the production and decay processes depend on the dark photon coupling to {electrons}, the presence of the dark dipole can affect both. 

The number of dark photons produced can be computed using
\be
N_{A'} = {\cal L} \sigma_{\rm prod}, 
\ee
where ${\cal L} \simeq 10^8$ pb$^{-1}$ is the effective luminosity\footnote{The luminosity is computed as $ \mathcal{L}=N_e X_\text{Al} N_A/A_\text{Al},
$ where $N_e=30$ C, $X_\text{Al}=23.4~\text{g/cm}^2$ is the Aluminium radiation length, $A_\text{Al}$ is its atomic mass and $N_A$ is Avogradro's number.} on the target and $\sigma_{\rm prod} = \sigma(e^- \,{\rm Al} \to A'\, e^-\, {\rm Al})$ is the Bremsstrahlung cross section. To compute it, we have adapted the {\tt Feynrules}~\cite{Alloul:2013bka} file of~\cite{1412.0018} to include the dipole interaction of Eq. \eqref{eq:Lint2}. Following~\cite{FormFactor}, we also implemented in {\tt Madgraph5}~\cite{Alwall:2014hca} the Aluminum form factor described in~\cite{Bjorken:2009mm}. The number of events of Eq.~\eqref{eq:Nevts} is computed using~{\tt MadDump}~\cite{1812.06771}, a module dedicated to beam dump experiments that takes care of computing not only $N_{A'}$, but $f_{\rm geom}$ and $P_{\rm dec}$ as well. We fix the parameters of the experiment according to~\cite{Bjorken:1988as}. We consider the decays $A' \to e^+e^-$ and $A' \to \mu^+ \mu^-$ in our analysis. To properly include the displaced decay of the dark photon in {\tt MadDump}, it is necessary to add the command \verb!add process displaced_decay Ap! (where \verb!Ap! stands for the dark photon) after the definition of the decay. For the purpose of validation we have cross-checked the results obtained from~\verb!MadDump! with our own implementation of the experiment, finding excellent agreement.
\begin{figure}[tb]
\centering
\includegraphics[width=1.0\textwidth]{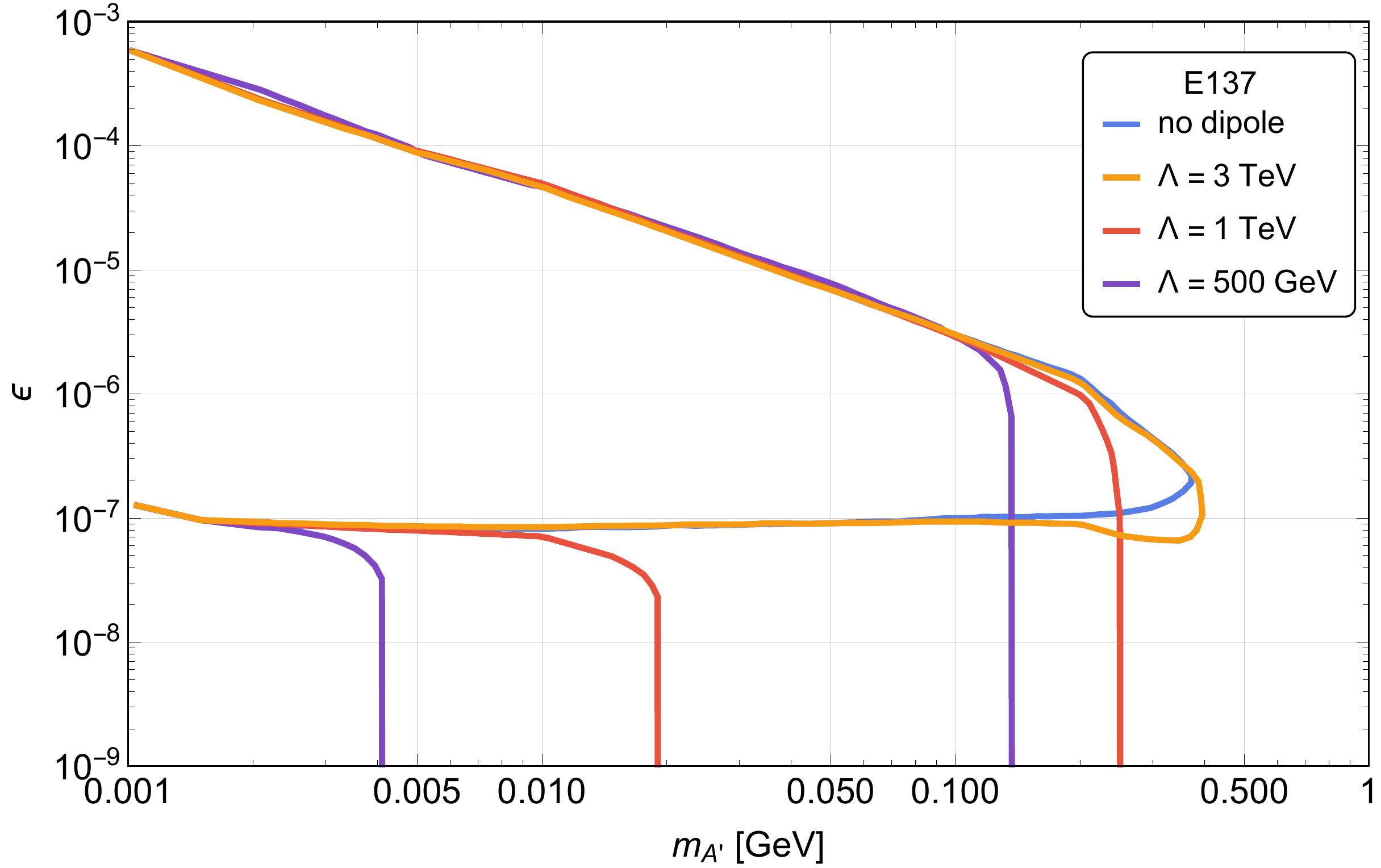}
\caption{Exclusion region obtained for E137 assuming the detection of at least 3 dark photon events. The blue curve is the limit obtained in the absence of the dipole operator, while the other curves show the limits for different values of $\Lambda$ by fixing ${d_e}=1$.}
\label{fig:E137_plot}
\end{figure}

To understand how the presence of the dark dipole affects the dark photon production and decay it is useful to inspect the modifications of the Bremsstrahlung production cross section and decay probability. This is shown in Fig.~\ref{fig:E137_xsec} for $\epsilon = 10^{-5}$ (dashed curves) and  $\epsilon = 10^{-8}$ (solid curves). The blue lines show the behavior for ${d_e} = 1$ and $\Lambda = 1$ TeV, while the red lines show the case without dipole. For the computation of the decay probability we have fixed a boost factor $\gamma = 4\times 10^3$. We have checked, using the numerical simulations, that this value is often obtained. For $\epsilon = 10^{-5}$ the presence of the dipole is practically irrelevant, confirming our previous claim that the effect of the dipole is very small in the large $\epsilon$ part of parameter space. On the contrary, for smaller values of $\epsilon$ (for instance, $\epsilon = 10^{-8}$ as shown in the figure) the effect of the dipole is important in both production and decays: the production cross section remains almost flat in the mass region of interest, while the decay probability becomes larger due to the larger value of the decay width. Combining the two effects, we obtain the results shown in Fig.~\ref{fig:E137_plot}. For $\Lambda$ smaller than $3$ TeV, the excluded region increases dramatically with respect to the no-dipole case, extending indefinitely to small values of $\epsilon$. This is because, for $\epsilon \lesssim 10^{-7}$, the kinetic mixing contribution is subdominant with respect to the dark dipole one and the number of events becomes independent of $\epsilon$. As in the LSND case, once we reach $\Lambda = 3$\;TeV the dipole effect is already almost decoupled, and only a small deformation of the exclusion region is obtained.

\section{Conclusions}\label{sec:conclusion}

How robust are the bounds on the visible dark photon? This is the question we have addressed in this work. Assuming for definiteness the existence of heavy NP interacting with both the SM and the dark photon, we have framed the problem in terms of a dark EFT. After a proper diagonalization of the kinetic and mass terms, we are left with all physical effects encoded in dipole operators. 

From the phenomenological side, in this paper we have focused on the consequences of a dark \textit{electron} dipole moment. We have analyzed its effect on BBN, CMB, supernov\ae\ and terrestrial bounds. Although we could naively expect the effect of heavy NP to give rather small deformations of the existing bounds, this is not always the case and depends strongly on the interplay between the value of $\Lambda$ and the value of the kinetic mixing parameter to which the bounds extend. The following picture emerges:
\begin{itemize}
    \item For kinetic mixing larger than $10^{-7}$ the effect of the dipole is too small to give any appreciable modification on the bounds;
    \item In the terrestrial experiments whose bounds extend down to $\epsilon \sim 10^{-7}$ (LSND and E137) the effect of the dipole decouples at scales around $\Lambda = 3$\;TeV. For smaller values of $\Lambda$ the effect is dramatic, up to the point that for E137 the excluded region is extended indefinitely to smaller values of $\epsilon$ in specific mass intervals, see Fig.~\ref{fig:LSND_plot} and Fig.~\ref{fig:E137_plot};
    \item All supernov\ae\ bounds become weaker, with effects that may again be dramatic (see Fig.~\ref{fig:SN_plot}). In this case, the effect of the dipole decouples at scales ${\cal O}(10\;{\rm TeV})$ and ${\cal O}(100\;{\rm TeV})$ for the luminosity and mantle arguments respectively.
    \item Finally, also the CMB and BBN bounds are drastically modified: they are ineffective even for heavy NP associated with scales of several tens of TeV. 
\end{itemize}
Altogether, we see that the effect of the dark {electron} dipole becomes more and more important the smaller the values of $\epsilon$ to which the experimental bound extends. Moreover, the smaller the kinetic mixing, the larger the scale $\Lambda$ at which the effect decouples. Our results are summarized in~Fig.~\ref{fig:all_bounds_intro}. {Our conclusions rest on a number of assumptions. First of all, we are supposing that the kinetic mixing and the electron dark dipole can be taken as independent parameters. As we have shown explicitly in Sec.~\ref{sec:UV_compl}, this may require some tuning in the parameters of the UV theory. Moreover, a further tuning may be needed to ensure $d_e \gg y_e$, where $y_e$ is the electron Yukawa coupling. If the UV completion is natural and such double-tuning is absent, we expect $\epsilon \sim d_e \sim y_e$. In this case the effect of the dipole would be too small to affect terrestrial and supernovae bounds but could still modify the BBN and CMB bounds.}

Our study may be extended in various directions. For instance, it would be interesting to study the physical effect of dark {\it quark} dipoles, or how the presence of a dark dipole would modify the limits of an {\it invisible} dark photons. We will come back to these aspects in future works.

\acknowledgments

G.G.d.C. thanks L. Darm\'e for useful discussions.
E.B. anknowledges financial support from ``Funda\c{c}\~ao de Amparo \`a Pesquisa do Estado de S\~ao Paulo'' (FAPESP) under contract 2019/04837-9.
G.G.d.C. is supported by the INFN Iniziativa Specifica Theoretical
Astroparticle Physics (TAsP) and by the Frascati National 
Laboratories (LNF) through a Cabibbo Fellowship call 2019. G.M.S. has received financial support from ``Fundação de Amparo à Pesquisa de São Paulo" (FAPESP) under contract 2020/14713-2.

\appendix

\section{Diagonalization of the kinetic and mass Lagrangians}\label{app:diagonalization}
In this Appendix we give more details on the diagonalization of the gauge kinetic and mass Lagrangian. Starting from Eq.~\eqref{eq:L_ren} and Eq.~\eqref{eq:L_EFT}, the kinetic Lagrangian reads
\al{\label{eq:Lkinetic}
{\cal L}_{kin} & = - \left( \frac{1}{4 \tilde{g}'^2} -  \frac{c_{BB} }{16\pi^2}\frac{v^2}{\Lambda^2} \right) B_{\mu\nu} B^{\mu\nu} -\left(  \frac{1}{4 \tilde{g}_d^2}  -  \frac{c_{XX}}{16\pi^2} \frac{v^2}{\Lambda^2} \right) X_{\mu\nu} X^{\mu\nu} \\
& ~~~~{} - \left( \frac{1}{4 \tilde{g}^2}- \frac{ c_{WW} }{16\pi^2}\frac{v^2}{\Lambda^2}  \right) W^3_{\mu\nu} W^{3\mu\nu} + \left(\frac{\kappa}{2 \tilde{g}' \tilde{g}_d}+ \frac{c_{XB}}{16\pi^2} \frac{v^2}{\Lambda^2}  \right) B_{\mu\nu} X^{\mu\nu} \\
& ~~~~ {} - \frac{c_{BW}}{32\pi^2} \frac{v^2}{\Lambda^2}  W^3_{\mu\nu} B^{\mu\nu} - \frac{c_{XW}}{32\pi^2} \frac{v^2}{\Lambda^2} W^3_{\mu\nu} X^{\mu\nu} \,  .
}
To diagonalize the kinetic terms, we first redefine the gauge couplings according to
\be
\tilde{g}_i^2 = \frac{g_i^2}{1+  \frac{c_i}{4\pi^2} \frac{v^2}{\Lambda^2} g_i^2} \, ,
\ee
for the appropriate Wilson coefficient. After these shifts, we define the usual kinetic mixing
\be\label{eq:epsilon}
\frac{\epsilon}{2 g' g_d}  \equiv \frac{\kappa \sqrt{1 +  \frac{c_{XX}}{4\pi^2} \frac{v^2}{\Lambda^2}  g_d^2} \sqrt{1+ \frac{c_{BB}}{4\pi^2} \frac{v^2}{\Lambda^2}  g'^2}}{2g' g_d}+  \frac{c_{XB}}{16\pi^2} \frac{v^2}{\Lambda^2} \, ,
\ee 
and the useful parameters
\be\label{eq:S_Sd}
{\cal S} = \frac{c_{BW}}{16\pi^2} \frac{ g g' \, v^2}{\Lambda^2} , ~~~~ {\cal S}_d = \frac{c_{XW}}{16\pi^2} \frac{g g_d\, v^2}{\Lambda^2} \, . 
\ee
By absorbing the gauge couplings in the gauge bosons fields we finally obtain
\be
{\cal L}_{kin} = -\frac{1}{4 } B_{\mu\nu} B^{\mu\nu} - \frac{1}{4} X_{\mu\nu} X^{\mu\nu} + \frac{\epsilon}{2} B_{\mu\nu}X^{\mu\nu} - \frac{{\cal S}}{2} W^3_{\mu\nu} B^{\mu\nu} - \frac{{\cal S}_d}{2} W^3_{\mu\nu} X^{\mu\nu} .
\ee
By defining the SM gauge bosons $Z_{\rm SM}$ and $A_{\rm SM}$ as the usual combination of the $B$ and $W^3$ weak bosons, the following field redefinitions allow to completely diagonalize the previous Lagrangian
\al{\label{eq:tr_can_kin_term}
& A_{{\rm SM}}  \to \left( 1- s_W c_W {\cal S} \right)A - {\cal S} c_{2W} \tilde{Z} + \left( c_W \epsilon - s_W {\cal S}_d \right) X \, ,\\
& Z_{{\rm SM}}  \to \left( 1 + s_W c_W {\cal S} \right) \tilde{Z} - \left(  c_W {\cal S}_d + \epsilon s_W \right) X \, .
}
The fields $A$, $\tilde{Z}$ and $X$ now have canonical kinetic terms. 
Let us now turn to the mass Lagrangian. Applying the shift just described, the mass matrix in the $(A, \tilde{Z}, X)$ basis results
\be
M^2 = 
\begin{pmatrix}
0 & 0 & 0 \\
0 & m_Z^2 \left( 1 +2 s_W c_W {\cal S} + {\cal T}\right) & - m_Z^2 \left( c_W {\cal S}_d + s_W \epsilon\right) \\ 
0 & - m_Z^2 \left(  c_W {\cal S}_d + s_W \epsilon\right)  & m_X^2
\end{pmatrix},
\ee
where we have defined 
\be
{\cal T} = \frac{c_T v^2}{\Lambda^2} \, ,
\ee
and $m_Z$ is the SM $Z$ boson mass. Since we are interested in a light dark photon we will take $m_X \ll m_Z$. To leading order in the NP parameters and in $m_X/m_Z$ we obtain a mixing angle
\be
\alpha \simeq s_W \epsilon + c_W {\cal S}_d ,
\ee
in such a way that
\al{
\tilde{Z} & \simeq Z + \left( s_W \epsilon + c_W {\cal S}_d\right) A'  , \\
X & \simeq A' - \left( s_W \epsilon + c_W {\cal S}_d \right) Z,
}
with $Z$ and $A'$ the mass eigenstates. The last step needed to obtain Eq.~\eqref{eq:Lint} is the usual electric charge redefinition needed when a $W^3_{\mu\nu} B^{\mu\nu}$ kinetic mixing is present:
\be
e \to \frac{e}{1 - s_W c_W {\cal S}}\, .
\ee
The same redefinitions of fields and parameters must be applied to the dipole operators. In full generality, we obtain the operators $\bar{\ell}_L \sigma^{\mu\nu} \ell_R A_{\mu\nu}'$, $\bar{\ell}_L \sigma^{\mu\nu} \ell_R A_{\mu\nu}$ and $\bar{\ell}_L \sigma^{\mu\nu} \ell_R Z_{\mu\nu}$ (plus the analogous involving quark fields), with Wilson coefficients given by combinations of the original coefficients and of the parameters entering the fields redefinitions. For simplicity, we do not write down these complete expressions.

\bibliographystyle{JHEP2}
\bibliography{Dark_photon}

\end{document}